\documentclass{llncs}

\usepackage{breakurl} 
\usepackage{makeidx}
\usepackage{graphicx,latexsym,alltt}
\usepackage[final]{hyperref}
\usepackage{amssymb,amsmath}
\usepackage{algorithm,algorithmic}
\usepackage{color}
\usepackage{url}
\usepackage{listings}
\usepackage{multirow} 
\usepackage{enumitem} 

\usepackage{multicol}
\usepackage{lipsum}

\usepackage{subcaption}

\usepackage{wrapfig}

\newcommand{\biblio}[1]{../../../../../../Biblio/#1}

\definecolor{blank}{rgb}{0.55,0.55,0.55}

\long\def\comment#1{}

\renewcommand{\phi}{\varphi}

\def\defemb#1#2{\expandafter\def\csname #1\endcsname
                              {\relax\ifmmode #2\else\hbox{$#2$}\fi}}
\defemb{cA}{{\cal A}}
\defemb{cB}{{\cal B}}
\defemb{cC}{{\cal C}}
\defemb{cD}{{\cal D}}
\defemb{cE}{{\cal E}}
\defemb{cF}{{\cal F}}
\defemb{cG}{{\cal G}}
\defemb{cH}{{\cal H}}
\defemb{cI}{{\cal I}}
\defemb{cJ}{{\cal J}}
\defemb{cL}{{\cal L}}
\defemb{cM}{{\cal M}}
\defemb{cN}{{\cal N}}
\defemb{cO}{{\cal O}}
\defemb{cP}{{\cal P}}
\defemb{cR}{{\cal R}}
\defemb{cS}{{\cal S}}
\defemb{cT}{{\cal T}}
\defemb{cU}{{\cal U}}
\defemb{cV}{{\cal V}}
\defemb{cW}{{\cal W}}
\defemb{cX}{{\cal X}}
\defemb{cZ}{{\cal Z}}

\makeatletter
\newenvironment{prog}{\vspace{1.0ex}\par
\obeylines\@vobeyspaces\tt}{\vspace{1.0ex}\noindent
}
\makeatother
\newcommand{\startprog}{\begin{prog}}
\newcommand{\stopprog}{\end{prog}\noindent}

\catcode`\_=\active
\let_=\sb
\catcode`\_=12
\makeatother

\newif\ifpaperVersion
\paperVersionfalse
\ifpaperVersion
	\newcommand{\ignore}[1]{}
	\newcommand{\deleted}[1]{}
	
	\newcommand{\pending}[1]{}
	\newcommand{\done}[1]{}
	\newcommand{\doubt}[1]{}
	\newcommand{\josep}[1]{}
	\newcommand{\david}[1]{}
	\newcommand{\sergio}[1]{}
	\newcommand{\tama}[1]{}
\else
	\definecolor{ignoreColor}{rgb}{1,0.5,0}
	\definecolor{pendingColor}{rgb}{0.2,0.7,0.2}
	\definecolor{doneColor}{rgb}{0.7,0.2,0.7}
	\definecolor{doubtColor}{rgb}{0.6,0.6,0.4}
	\definecolor{josepColor}{rgb}{0.2,0.6,0.6}
	\definecolor{davidColor}{rgb}{0.6,0.2,0.6}
	\definecolor{sergioColor}{rgb}{1.0,0.5,0.0}
 	\definecolor{tamaColor}{rgb}{0.2,0.8,1.0}

	\newcommand{\ignore}[1]{\textcolor{ignoreColor}{\#Ignored: #1}}
	\newcommand{\deleted}[1]{\textcolor{red}{\#Deleted: #1}}
	
	\newcommand{\pending}[1]{\textcolor{pendingColor}{\#Pending: #1}}
	\newcommand{\done}[1]{\textcolor{doneColor}{\#Done: #1}}
	\newcommand{\doubt}[1]{\textcolor{doubtColor}{\#Doubt: #1}}
	\newcommand{\josep}[1]{\textcolor{josepColor}{\#JJJ: #1}}
	\newcommand{\david}[1]{\textcolor{davidColor}{\#DDD: #1}}
	\newcommand{\sergio}[1]{\textcolor{sergioColor}{\#SSS: #1}}
	\newcommand{\tama}[1]{\textcolor{tamaColor}{\#TTT: #1}}
\fi

\usepackage{listings}
\usepackage[utf8]{inputenc}
\usepackage{courier}
\usepackage[T1]{fontenc}
\usepackage[scaled]{couriers}

\usepackage[utf8]{inputenc}
\usepackage[T1]{fontenc}
\usepackage[svgnames]{xcolor}
\usepackage[scaled=0.84]{beramono}
\usepackage{listings}

\makeatletter
\newcommand{\MyChange}[2]{%
  \extractcolorspec{.}\MyChange@CurrentColor
  \extractcolorspec{#2}\MyChange@TestColor
  \ifx\MyChange@CurrentColor\MyChange@TestColor
    {\bfseries\color{green!33!black} #1 }%
  \else
    {#1 }%
  \fi%
}
\makeatother

\lstdefinelanguage{Erlang}
{
classoffset=1,
keywords={},
  keywords = [2]{receive, if, case, try, module, export, all, spawn, end, fun, when, after, spec, of,  compile, true, false, assertEqual, assertError, include, lib},
  keywordstyle=\textbf,
  keywordstyle=[2]\color{blue},
classoffset=2,
  otherkeywords={
    ;, ",\,,\#, ., (, ), \{, \}, [,],|,||,+,-,*,/, <,>, >=, =<,->, =, ==, =:=, _,?
  },
  classoffset=0,
  sensitive=true,
   morecomment=[l]{\%},
     commentstyle=\color{red}\emph,
  stringstyle=\color{black},
  morestring=[b]',
}

\lstdefinelanguage{none}
{
keywords={},
keywords = [2]{},
otherkeywords={},
morekeywords={}
  ,classoffset=1
  ,sensitive=true
  ,stringstyle=\color{black}
  ,morestring=[b]',
  numbers=none,
  ,literate=
}

\begin{document}

\frontmatter
\pagestyle{headings}
\addtocmark{WFLP}

\pagenumbering{arabic}

\title
{
Runtime verification in Erlang by using contracts
 \thanks
 {
 This work has been partially supported by MINECO/AEI/FEDER (EU)
 under grant TIN2016-76843-C4-1-R, by
 the \emph{Comunidad de Madrid} under grant S2013/ICE-2731
 (\emph{N-Greens Software}),
and by the \emph{Generalitat Valenciana} under grant
PROMETEO-II/2015/013 (\emph{SmartLogic}).
Salvador Tamarit was partially supported by the \emph{Conselleria de
Educaci\'on, Investigaci\'on, Cultura y Deporte de la Generalitat
Valenciana} under grant APOSTD/2016/036.
  }
}
\titlerunning{Runtime verification in Erlang by using contracts}

\author{Lars-\AA ke Fredlund\inst{1} \and Julio Mariño\inst{1} \and
Sergio Pérez\inst{2} \and Salvador Tamarit\inst{2}}

\institute{Babel Group\\
Universidad Polit\'ecnica de Madrid, Spain\\
\email{\{lfredlund,jmarino\}@fi.upm.es}
\and
Departament de Sistemes Inform\`atics i Computaci\'o\\
Universitat Polit\`ecnica de Val\`encia, Spain\\
\email{\{serperu, stamarit\}@dsic.upv.es}}

\maketitle


\begin{abstract}

During its lifetime, a program suffers several changes that seek to improve or to augment some parts of its functionality. However, these modifications usually also introduce errors that affect the already-working code. There are several approaches and tools that help to spot and find the source of these errors. However, most of these errors could be avoided beforehand by using some of the knowledge that the programmers had when they were writing the code. This is the idea behind the design-by-contract approach, where users can define contracts that can be checked during runtime. In this paper, we apply the principles of this approach to Erlang, enabling, in this way, a runtime verification system in this language. We define two types of contracts. One of them can be used in any Erlang program, while the second type is intended to be used only in concurrent programs. We provide the details of the implementation of both types of contracts. Moreover, we provide an extensive explanation of each contract as well as examples of their usage. All the ideas presented in this paper have been implemented in a contract-based runtime verification system named \texttt{EDBC}. Its source code is available at GitHub as an open-source and free project.

\end{abstract}


\setlength{\columnsep}{1cm}

\begin{lstlisting}[float, language=none]
\end{lstlisting}

\section{Introduction}
\label{sec:intro}


Developing software is not an easy task and, consequently, some errors (\emph{bugs}) can be introduced in the code. These errors can be really simple to solve but not so easy to find. Companies usually rely on program testing to check that their developed programs behave as expected. There are also programming languages that provide a robust static type systems which can unveil some errors during program compilation. These or similar methods are really useful to find the symptom of a bug. Once a symptom is found,  the debugging process can start in order to find the source of the bug. The cause can be in a place that goes from the same line reported by the error to other modules. Most of the bug sources would be found easier and sooner if users defined further control in their code. For instance, consider a function which can perform a division by zero if one of its parameters has a non-expected value. Without any special control, the program would fail and report a division by zero if one of the non-expected values were used in the call. However, the bug source is in the call and not in the reported division operation. Defensive programming is a way to avoid non-expected values which produce unexpected behaviours, e.g. divisions by zero. The idea is to check the validity of the arguments before doing the real job. However, defensive programming is not a recomendable practice, for various reasons, like the boiler-plate or the overhead introduced in the final release. In fact, in our running language, Erlang, this practice is always banned in the good-programmer manuals. 

Erlang innately has a lot of ways to control how to recover from errors, but these features are supposed to be used only for plausible errors. Unfortunatelly, not all the errors fall in this category.  There are errors that simply should never happen, and they should be fixed during the software production. Erlang, as Python or JavaScript, is a dynamic programming language. This means that there is not a really strict control during compilation that could help avoiding errors before runtime. For this reason, static analysis techniques, led by Dialyzer \cite{LinS04}, have been successfully and widely adopted by Erlang's practitioners during the last years. Dialyzer can analyze the code and report some errors without any special input from users. However, its results can be considerably improved by the use of type contracts \cite{JLS07}. These contracts are not there to be only used by Dialyzer or similar tools, but also as a documentation that could improve the maintainability of the resulting software.

Dialyzer and testing, e.g. EUnit, are common Erlang tools used to find errors before the final software deployment. However, there are still some errors that can escape from the claws of these tools. Unfortunately, these errors can appear when the program is being used by the final user. In most of the cases, these errors have their source in a knowledge that users had when they were writing the code, e.g. that the result of a function must be greater than the first parameter. Unfortunately, programming languages rarely provide a method to input this information beyond defensive programming. In this work, we propose an alternative to reflect this knowledge: the Erlang Design-By-Contract (\texttt{EDBC}) system, a runtime verification system based on the Design-By-Contract \cite{Mey92} philosophy. The \texttt{EDBC} system is available as free and open-source software in \texttt{GitHub}: \url{https://github.com/tamarit/edbc}. 

The Design-By-Contract approach is based on the definition of contracts which are checked for validity during runtime. Then, this approach allows users to add extra knowledge which can help to detect the bug source more easily and sooner. Unfortunately, contract checking usually comes with an overhead in the runtime. However, the contract checking is meant to be done only during software production, and it should be disabled when the final release is generated. Once disabled, the defined contracts can still be used as program documentation.

There are different types of contracts, most of them are related to program functions. The most common contracts are the preconditions and the postconditions. Preconditions are conditions that should hold before evaluating a function, while postconditions should hold after its evaluation. There are more types of contracts, e.g. the aforementioned type contracts (\texttt{spec}s in Erlang). In concrete, our approach includes pre and postcondition contracts, type contracts, decreasing-argument contracts, execution-time contracts and purity contracts. All these contracts can be used in any Erlang program, i.e. with a sequential or concurrent evaluation. 
The descriptions of the contracts that our system propose are in Section \ref{sec:contracts_desc}. We also provide some examples of how the contracts are used and what are the result of using them. These examples can be found in Section \ref{sec:contracts_exa}. Implementation details are also presented in the paper. In concrete, the contract checking is performed by a program instrumentation (described in Section \ref{sec:contracts_impl}). There are a second type of contracts which only can be used in concurrent evaluations, in concrete, in programs using Erlang behaviours. The Erlang behaviours are formalizations of common programming patterns. The idea is to divide the code for a process in a generic part (a behaviour module) and a specific part (a callback module). Thus, the behaviour module is part of Erlang/OTP and the callback is implemented by the user. An example of an Erlang behaviour is the \texttt{gen_server} behaviour, that implements the the generic parts of a server. Therefore, a callback module implementing this behaviour needs to define the specific parts of a server, e.g. how the state is initialized (\texttt{init/0}), how to serve a request (synchronous \texttt{handle_call/3} and  asynchronous \texttt{handle_cast/2}), etc. Contracts for this Erlang behaviours are described in Section \ref{sec:gen_server_cpre} while some examples of their usefulness can be found in Sections \ref{sec:sel_recv}-\ref{sec:rw}.
Finally, we discuss where our approach is placed among similar approaches in Section \ref{sec:related} before concluding the paper in Section \ref{sec:conclusions}.

\section{Contracts in Erlang}
\label{sec:contracts}




In this section we first introduce all the contracts available at the \texttt{EDBC} system (Section \ref{sec:contracts_desc}), then we show how they can be used to perform program verification (Section \ref{sec:contracts_exa}), and finally, we present some details of their implementation (Section \ref{sec:contracts_impl}).

\begin{itemize}

\subsection{The contracts}
\label{sec:contracts_desc}

\item \textbf{Precondition contracts}. With the macro \texttt{?PRE/1} we can define a precondition that a function should hold. The macro should be placed before the first clause of the contracted function. The single argument of this macro is a function without parameters, e.g. \texttt{fun  pre/0} or an anonymous function \texttt{fun() -> $\ldots$ end}, that we call \emph{precondition function}. A precondition function is a plain Erlang function. Its only particularity is that it includes references to the function parameters. In Erlang, a function can have more than one clause, so referring the parameter using the user-defined names can be confusing for both, for \texttt{EDBC} and also for the user. In order to avoid these problems, in \texttt{EDBC} we refer to the parameters by their position. Additionally, the parameters are rarely single variables and they can be more complex terms like a list or a tuple. We use the \texttt{EDBC}'s macro \texttt{?P/1} to refer to the parameters. The argument of this macro should be a number that ranges from 1 to the contracted function's arity. For instance, \texttt{?P(2)} refers to the second parameter of the function. A precondition function should return a boolean value which indicates whether the precondition of the contracted function is held or not. The precondition is checked during runtime before actually performing the call. Then, if the precondition function evaluates to \texttt{false}, the call is not preformed and a runtime error is raised.
\\

\item \textbf{Postcondition contracts}. In a similar way than in preconditions, we can use the macro \texttt{?POST/1} to define a postcondition that a function should hold. The macro should be placed after the last clause of the contracted function. Its argument is again a function without parameters, that we call \emph{postcondition function}. In this case, we need to refer to the result of the function, however Erlang has not any way to refer to it. Therefore, we use the \texttt{EDBC}'s macro \texttt{?R} to this purpose. Additionally, we can also use \texttt{?P/1} macros to refer to the contracted function's parameters within a postcondition function. The result of a postcondition function is also a boolean value. Therefore, a postcondition function is exactly the same as a precondition function with the only difference that a postcondition function can use the macro \texttt{?R} in its body. Postcondition functions are checked right after the call is completely evaluated and a runtime error is raised if they evaluate to \texttt{false}.
\\

\item \textbf{Decreasing-argument contracts}. These contracts are meant to be used in recursive functions and check that some of their arguments are always decreasing in each nested call. There are two types of macros to define these contracts: \texttt{?DECREASE/1} and \texttt{?SDECREASE/1}. They both operate exactly in the same way with the exception that the \texttt{?SDECREASE/1} macro indicates that the argument should be strictly lower in each nested call, while the \texttt{?DECREASE/1} macro also allows the argument to be equal. The argument of both macros can be either a single \texttt{?P/1} macro or a list containing several \texttt{?P/1} macros. These contracts should be placed before the first clause of the function. Decreasing-argument contracts are checked each time a recursive function is found, by comparing the arguments of the current call with the nested call just before performing the actual recursive call. In case the contracted decreasing argument is not actually decreasing, a runtime error is raised and the call is not performed. 
\\

\item \textbf{Execution-time contracts}. \texttt{EDBC} introduces two macros that allow users to define contracts related with execution time: \texttt{?EXPECTED_TIME/1}  and \texttt{?TIMEOUT/1}. The macros should be placed before the first clause of the contracted function. The argument of these macros is a function without parameters called \emph{execution-time function}. An execution-time function should evaluate to an integer which defines the expected execution time in milliseconds. Within the body of an execution-time function we can use \texttt{?P/1} macros to refer to the arguments. This feature is specially useful when dealing with lists or similar structures where the expected execution-time of the contracted function is related to their sizes. Both macros have a similar semantics, the only difference is that with macro \texttt{?EXPECTED_TIME/1} the \texttt{EDBC} system waits till the evaluation of the call finishes to check whether the call takes the expected time, while with macro \texttt{?TIMEOUT/1} \texttt{EDBC} raises an exception when the defined timeout is reached.
\\

\item \textbf{Purity contracts}. This contract allows users to declare that certain functions do not perform any impure operation. This purity condition of a contracted function can be declared by using the macro \texttt{?PURE/0} before its first clause. The purity checking process is performed in two steps. First, before a call to a contracted function is performed, a tracing process is started. Then, once the evaluation of the contracted function call finishes and if a call to an impure function or operation has been found during its evaluation, a runtime error is raised. The way we check the purity of a function ensures no false positives neither negatives.
The purity checking is not compatible with execution-time contracts which innately requiere to perform impure actions. 
\\

\item \textbf{Invariant contracts}. This contract is meant to be used in those Erlang behaviours with an internal and live state. An invariant contract is defined by using the macro \texttt{?INVARIANT/1}. This macro can be placed anywhere inside the module implementing the behaviour. The argument of the \texttt{?INVARIANT/1} macro is a function, named \emph{invariant function}, with only one parameter that represents the current state of the behaviour. Then, an invariant function should evaluate to a boolean value which indicates whether the given state satisfies the invariant or not. The invariant function is used to check the invariant contract each time a call to a state-modifier function , e.g. \texttt{handle_call/3} from the \texttt{gen_server} behaviour, finishes. Invariant contracts together with an enhanced state, can be used to control some relevant question as starvation. Examples of invariant contracts are presented in Section \ref{sec:gen_server_cpre}. 
\\

\item \textbf{Type contracts}. Erlang has a dynamic type system, i.e. the types are not checked during compilation. However, we still can define type contracts (represented by \texttt{spec} attributes) which serves for both, to document the code and to help static analyzers like \texttt{Dialyzer}\cite{LinS04}. Therefore, they are considered a powerful tool to anticipate type errors palliating in this way the main inconvenience of the dynamic typing approach. These type contracts are not dynamically checked by the Erlang interpreter and, of course, it has sense because these checks would have a cost which is not desirable when the software is already released. However, when the software is still in production, checking these type contracts during runtime can be very helpful to uncover unexpected behaviours. For this reason, before a function is evaluated, \texttt{EDBC} checks the type contract of its parameters (if any), while its result is checked after its evaluation. When some of the expected types, for the parameters or for the results, do not correspond with the obtained value, a runtime error is raised. Note that \texttt{EDBC} does not use any special macro to check type contracts, \texttt{spec} attributes are used instead.

\end{itemize}

\subsection{Verification of Erlang programs using contracts}
\label{sec:contracts_exa}

\begin{wrapfigure}[6]{r}{0.45\textwidth}
\vspace{-35pt}
\begin{center}
\begin{lstlisting}[basicstyle=\ttfamily\scriptsize, frame=single, 
%caption={Comparison function which ignores the callee}, label=lst:fc_ign_callee, 
language=erlang,
numbers=left, stepnumber=1,escapechar=@]
?PRE(fun() -> ?P(1) >= 0 end).
?SDECREASES(?P(1)).
fib(0) -> 0;
fib(1) -> 1;@\label{fig:fib_pre}@
fib(N) -> fib(N - 1) +  fib(N - 2).@\label{fig:fib_decr}@
\end{lstlisting}
\vspace{-20pt}
\end{center}
\caption{Fibonacci numbers}
\label{fig:fib}
\vspace{-10pt}
\end{wrapfigure}

In this section we show how to use the contracts presented in the previous section and also what are the benefits of using them.

We start by a simple example: the Fibonacci numbers. Figure \ref{fig:fib} shows an implementation in Erlang with some contracts attached. In concrete, the contracts are defining that the parameter should be a non-negative integer and that it should be strictly decreased in each recursive call.

\begin{wrapfigure}[5]{r}{0.5\textwidth}
\vspace{-30pt}
\begin{center}
\begin{lstlisting}[basicstyle=\ttfamily\scriptsize, frame=single,
% caption={\texttt{SecEr} reports that no discrepancies exist}, label=lst:happycorrect, 
language=none,escapechar=@]
** exception error: {"Decreasing condition does 
not hold. Previous call: fib(2). 
Current call: fib(4).", [{ex,fib,1,[]},@\ldots@
\end{lstlisting}
\vspace{-20pt}
\end{center}
\caption{Decrease contract violation}
\label{fig:fib_fail}
\vspace{-10pt}
\end{wrapfigure}

In case any of the contracts is found to be false, an error is raised. For instance, if we change \texttt{fib(N - 2)} in line \ref{fig:fib_decr} to \texttt{fib(N + 2)}, the contract-violation error shown in Figure \ref{fig:fib_fail} is raised.

\begin{wrapfigure}[7]{r}{0.65\textwidth}
\vspace{-30pt}
\begin{center}
\begin{lstlisting}[basicstyle=\ttfamily\scriptsize, frame=single,
% caption={\texttt{SecEr} reports that no discrepancies exist}, label=lst:happycorrect, 
language=none,escapechar=@]
method Find(a: array<int>, key: int) returns (index: int)
   requires a != null
   ensures 0 <= index ==> index < a.Length && a[index] == key
   ensures index < 0 ==> forall k :: 
      0 <= k < a.Length ==> a[k] != key
{@\ldots@}
\end{lstlisting}
\vspace{-20pt}
\end{center}
\caption{Contracted function \texttt{Find/2} in \texttt{Dafny}}
\label{fig:find_dafny}
\vspace{-10pt}
\end{wrapfigure}

We can use the \texttt{EDBC} system to define more advanced contracts. For instance, \texttt{Dafny} \cite{Lei10}, which was an inspiration for our system, 
introduces the quantifiers as a way to define conditions for input lists. Figure \ref{fig:find_dafny} shows how these quantifiers are declared in \texttt{Dafny} for a function \texttt{Find/2} which simply search for the position of a value in an array and return \texttt{-1} if it is not found. 

\begin{wrapfigure}[8]{r}{0.55\textwidth}
\vspace{-10pt}
\begin{center}
\begin{lstlisting}[basicstyle=\ttfamily\scriptsize, frame=single, 
language=erlang,
numbers=left, stepnumber=1,escapechar=@]
?PRE(fun() -> length(?P(1)) > 0 end).
find(L, K) -> @\ldots@
?POST(fun() -> ?R < 0 orelse  
        (?R < length(?P(1)) 
         andalso lists:nth(?R, ?P(1)) == ?P(2)) end).
?POST(fun() -> ?R > 0 orelse 
        lists:all(fun(K) -> K /= ?P(2) end, ?P(1)) end).
\end{lstlisting}
\vspace{-20pt}
\end{center}
\caption{Contracted function \texttt{find/2} in \texttt{Erlang}}
\label{fig:find_erlang}
\vspace{-10pt}
\end{wrapfigure}

These contracts with quantifiers can be also declared in Erlang using our approach. Instead of using a special syntax like in \texttt{Dafny}, we can check conditions with quantifiers using common Erlang function like \texttt{lists:all/2} which checks whether a given predicate is true for all the elements of a given list. Figure \ref{fig:find_erlang} shows how the contracts in Figure \ref{fig:find_dafny} are implemented in Erlang. If we implemented this function as a recursive function the list would be decreasing between recursive calls. Then, we could also add the contract \texttt{?SDECREASE(?P(1))} to the function.

\begin{wrapfigure}[18]{r}{0.5\textwidth}
\vspace{-30pt}
\begin{center}
\includegraphics[width=0.5\textwidth]{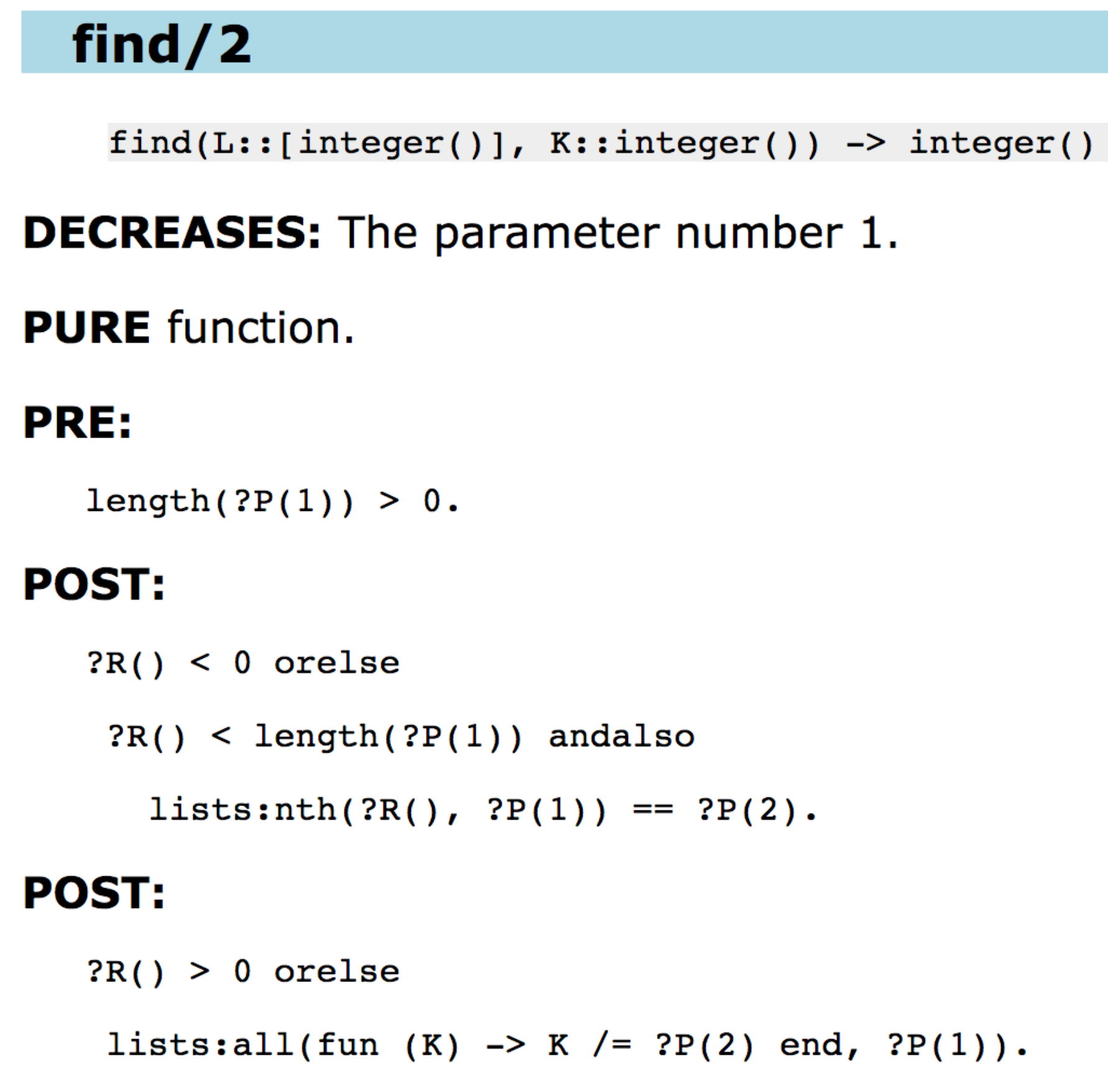}
\vspace{-20pt}
\end{center}
\caption{\texttt{EDoc} for the contracted function \texttt{find/2}}
\label{fig:find_doc}
\vspace{-10pt}
\end{wrapfigure}

Note that two postconditions are defined for function \texttt{find/2} in Figure \ref{fig:find_erlang}. In this function it is pretty clear which one failed by observing the result, i.e. if it is negative or not, but it could happen that for other contracts it is not so clear. We provide an alternative output for the contract functions that allows to identify the failing contract more easily. Instead of defining a contract function that returns boolean values, users can also define a function that returns tuple containing a boolean value and as second element a message to be shown when the condition fails. This can be useful not only to distinguish among several contracts, but also to attach additional information which can help to easily identify the source of the unexpected behaviour. Additionally, users can add some logging data in all type of contracts. In this way, users can trace some relevant information during execution.

\begin{wrapfigure}[6]{r}{0.55\textwidth}
\vspace{-35pt}
\begin{center}
\begin{lstlisting}[basicstyle=\ttfamily\scriptsize, frame=single, 
language=erlang,
numbers=left, stepnumber=1,escapechar=@]
fold1(Fun, Acc, Lst) -> lists:fold(Fun, Acc, Lst).
fold2(Lst, Fun) -> fold1(Fun, 1, Lst).
g3() -> fold1(fun erlang:put/2, ok, [computer, error]).
?PURE.
g4() -> fold2([2, 3, 7], fun erlang:'*'/2).
\end{lstlisting}
\vspace{-20pt}
\end{center}
\caption{Example taken from \texttt{PURITY} \cite{PS10}}
\label{fig:fold_pure}
\vspace{-10pt}
\end{wrapfigure}

Contracts added by users can also be used to generate documentation. Erlang OTP includes the tool \texttt{EDoc} \cite{edoc} which generates  documentation for modules in \texttt{HTML} format. We have modified the resulting \texttt{HTML} to add information of contracts. An example of an \texttt{EDoc}-generated documentation for function \texttt{find/2} with information of its contracts (some in Figure \ref{fig:find_erlang} and some new) and its \texttt{spec}s is depicted in Figure \ref{fig:find_doc}.

\begin{wrapfigure}[6]{r}{0.47\textwidth}
\vspace{-10pt}
\begin{center}
\begin{lstlisting}[basicstyle=\ttfamily\scriptsize, frame=single,
% caption={\texttt{SecEr} reports that no discrepancies exist}, label=lst:happycorrect, 
language=none,escapechar=@]
** exception error: {"The function is not pure. 
Last call: ex:g3(). 
It has call the impure BIF erlang:put/2 
when evaluating g3().",[{ex,g3,0,[]}]}
\end{lstlisting}
\vspace{-20pt}
\end{center}
\caption{Pure contract-violation report}
\label{fig:fold_pure_error}
\vspace{-10pt}
\end{wrapfigure}

In order to show purity contracts we take a simple example used to present \texttt{PURITY} \cite{PS10}, an analysis that statically decides whether a function is pure or not. Although our goal here is different, it is still interesting to study what happens in the cases that they found more problematic. In concrete, its analysis loses some accuracy in cases where (in their own words) \emph{multiple levels of indirection are present between a call with a concrete function as argument and the actual higher order function which will end up using it}. 

\begin{wrapfigure}[8]{r}{0.55\textwidth}
\vspace{-30pt}
\begin{center}
\begin{lstlisting}[basicstyle=\ttfamily\scriptsize, frame=single, 
language=erlang,
numbers=left, stepnumber=1,escapechar=@]
?EXPECTED_TIME(fun() ->
    20 + lists:sum([case (I rem 2) of
    0 -> 100; 1 -> 200 end|| I <- L]) end)
f_time(L) -> [f_time_run(E) || E <- L].
f_time_run(N) when (N rem 2) == 0 ->  timer:sleep(100);
f_time_run(N) when (N rem 2) /= 0 -> timer:sleep(200).
\end{lstlisting}
\vspace{-20pt}
\end{center}
\caption{Function with time contracts}
\label{fig:time}
\vspace{-10pt}
\end{wrapfigure}

The example the authors presented is depicted in Figure \ref{fig:fold_pure}. We only added the contract \texttt{?PURE} in the test case \texttt{g4/0}, because the other test case, i.e. \texttt{g3/0}, performs the impure operation \texttt{erlang:put/2}. When \texttt{g4/0} is run, no contract violation is reported as expected. If we added the contract \texttt{?PURE} to \texttt{g3/0}, then the execution will fail with the error shown in Figure \ref{fig:fold_pure_error}.

\begin{wrapfigure}[5]{r}{0.62\textwidth}
\vspace{-30pt}
\begin{center}
\begin{lstlisting}[basicstyle=\ttfamily\scriptsize, frame=single,
% caption={\texttt{SecEr} reports that no discrepancies exist}, label=lst:happycorrect, 
language=none,escapechar=@]
** exception error: {"The execution of 
ex:f_time2([1,2,3,4,5,6,7,8,9,10]) took too much time.
Real: 1509.913 ms. Expected: 1020 ms. Difference: 489.913 ms)
\end{lstlisting}
\vspace{-20pt}
\end{center}
\caption{Execution-time contract-violation report}
\label{fig:time_error}
\vspace{-10pt}
\end{wrapfigure}

In order to understand how the time contracts work, we present an example of a function that performs a list of tasks. Each task has its type (even or odd), and the time is defined by this type (100 and 200 ms., respectively). Figure \ref{fig:time} shows this function and its associated time contract. In case we changed the execution-time function to something like \texttt{fun() -> 20 + (length(?P(1)) * 100) end}, we would obtain the contract-violation report shown in Figure \ref{fig:time_error}.


\begin{wrapfigure}[5]{r}{0.6\textwidth}
\vspace{-30pt}
\begin{center}
\begin{lstlisting}[basicstyle=\ttfamily\scriptsize, frame=single,
% caption={\texttt{SecEr} reports that no discrepancies exist}, label=lst:happycorrect, 
language=none,escapechar=@]
** exception error: {"The spec precondition does not hold.
Last call: ex:fib(a).  
The value a is not of type integer().", @\ldots@}
\end{lstlisting}
\vspace{-20pt}
\end{center}
\caption{\texttt{spec} contract-violation report}
\label{fig:spec_error}
\vspace{-10pt}
\end{wrapfigure}

For \texttt{spec} checking, users only have to define the function \texttt{spec}s as usual or reuse the existing ones. Then, they run the code and an error will be raised only in case that some of the arguments or the result of a function call have an unexpected type. Figure \ref{fig:spec_error} shows an error that would be shown in case of calling \texttt{fib(a)} when the \texttt{spec} \texttt{fib(integer()) -> integer()} is defined in function \texttt{fib/1}. Finally, we want to highlight that the contract checking performed by \texttt{EDBC} has not any incompatibility with other tools. For instance, users can define contracts and also EUnit \cite{CR06} assertions in the same function. All these and more examples are collected at the \texttt{GitHub} repository\footnote{\url{https://github.com/tamarit/edbc/tree/master/examples}}. 


\subsection{Implementation details}
\label{sec:contracts_impl}

In this section we explain how we are instrumenting the code to get contract checking during runtime. Note that the code produced by the instrumentation process is normal Erlang code meant to be run in the standard interpreter of the language. Given a module,
we run the algorithm in Figure \ref{fig:inst_alg} for each of its function definitions. The result of this algorithm is a collection of function definitions which contains the (maybe modified) input function definition and some helper functions which are synthetized according to the contracts. The instrumentation is performed in three steps. 

\begin{figure}[!t]
\centering
\begin{subfigure}{.45\textwidth}
\begin{center}
\begin{lstlisting}[basicstyle=\ttfamily\tiny, frame=single, 
%caption={Comparison function which ignores the callee}, label=lst:fc_ign_callee, 
language=none,
numbers=left, stepnumber=1,
escapechar=@]
@\textbf{INPUT:}@ A function definition @$\mathtt{f_{def} = \overline{f_{name}(\overline{p_n}) \rightarrow \overline{e_m}._{_o}}}$@
@\textbf{OUTPUT:}@ A list of function definitions

contracts = read_contracts(@$\mathtt{f_{def}}$@)
funs = @$\mathtt{\emptyset}$@
if @$\mathtt{contracts \neq \emptyset} $@ @\label{fig:inst_alg_pi}@
    (@$\mathtt{f_{def}}$@, funs) = inst_put_info(@$\mathtt{f_{def}}$@, funs, n) @\label{fig:inst_alg_pi_2}@
if @$\mathtt{\exists c \in contracts . type(c) = decrease} $@ @\label{fig:inst_alg_decr}@
    (@$\mathtt{f_{def}}$@, funs) = inst_decr(c, @$\mathtt{f_{def}}$@, @$\mathtt{f_{name}}$@, funs, n) 
    contracts = contracts \ {c}
foreach @$\mathtt{c \in contracts}$@
    if  type(c) = pre
        (@$\mathtt{f_{def}}$@, funs) = inst_pre(c, @$\mathtt{f_{def}}$@, funs, n)
    else
        (@$\mathtt{f_{def}}$@, funs) = inst_post(c, @$\mathtt{f_{def}}$@, funs, n)     
    contracts = contracts \ {c}
return @$\mathtt{funs \cup \{f_{def}\}}$@
\end{lstlisting}
\vspace{-20pt}
\end{center}
\caption{Instrumentation of a contracted function}
\label{fig:inst_alg}
\end{subfigure}%
~~~~~~~~~
\begin{subfigure}{.45\textwidth}
\begin{center}
\begin{lstlisting}[basicstyle=\ttfamily\tiny, frame=single, 
%caption={Comparison function which ignores the callee}, label=lst:fc_ign_callee, 
language=none,
numbers=left, stepnumber=1,
escapechar=@]
inst_decr(@$\mathtt{?DECREASE(\overline{?P_o})}$@, @$\mathtt{f_{def}}$@, @$\mathtt{f_{original\_name}}$@, funs, n) =
    @$\mathtt{f'_{name}}$@ = get_free_name()
    @$\mathtt{f_{name}}$@ = get_name(@$\mathtt{f_{def}}$@)
    @$\mathtt{nfuns}$@ = {@$\mathtt{f'_{name}}$@@$\mathtt{(\overline{fv'_o}, \overline{fv_n}) \rightarrow}$@ edbc_lib:decrease_check( @\label{fig:inst_decr_new}@
        @$\mathtt{\overline{fv'_o}, \overline{fv_n}|_{\overline{?P_o}}}$@,
        fun() @$\mathtt{\rightarrow}$@ @$\mathtt{f_{original\_name}}$@@$\mathtt{(\overline{fv_n})}$@ end).}
    @$\mathtt{f_{def}}$@ = 
        @$\mathtt{\{f_{name}(\overline{p_n}) \rightarrow }$@ @\label{fig:inst_decr_replace1}@
            @$\mathtt{replace(body, \{f_{name}(\overline{e_n}) \mapsto f'_{name}(\overline{p_n}|_{\overline{?P_o}}, \overline{e_n})\})}$@ @\label{fig:inst_decr_replace2}@
            @$\mathtt{|~ f_{name}(\overline{p_n}) \rightarrow body \in f_{def}\}}$@
    return (@$\mathtt{f_{def}}$@, funs @$\cup$@ @$\mathtt{nfuns}$@)
\end{lstlisting}
\vspace{-20pt}
\end{center}
\caption{Instrumentation of a \texttt{?DECREASE} contract}
\label{fig:inst_decr}
\end{subfigure}
~~~~~~~~~
\begin{subfigure}{.45\textwidth}
\begin{flushright}
\begin{lstlisting}[basicstyle=\ttfamily\tiny, frame=single, 
%caption={Comparison function which ignores the callee}, label=lst:fc_ign_callee, 
language=none,
numbers=left, stepnumber=1,
escapechar=@]
inst_pre(@$\mathtt{?PRE(f_c)}$@, @$\mathtt{f_{def}}$@, funs, n) =
    @$\mathtt{f'_{name}}$@ = get_free_name()
    @$\mathtt{f_{name}}$@ = get_name(@$\mathtt{f_{def}}$@)
    @$\mathtt{nfuns}$@ = {@$\mathtt{f_{name}}$@@$\mathtt{(\overline{fv_n}) \rightarrow}$@ edbc_lib:pre(
        fun() @$\mathtt{\rightarrow}$@ replace(@$\mathtt{body(f_c), \{?P(i) \mapsto fv_i\}}$@) end, 
        fun() @$\mathtt{\rightarrow}$@ @$\mathtt{f'_{name}}$@@$\mathtt{(\overline{fv_n})}$@ end).}
    return (rename_fun(@$\mathtt{f_{def}}$@, @$\mathtt{f'_{name}}$@), funs @$\cup$@ @$\mathtt{nfuns}$@)
\end{lstlisting}
\vspace{-20pt}
\end{flushright}
\caption{Instrumentation of a \texttt{?PRE} contract}
\label{fig:inst_pre}
\end{subfigure}
~~~~~~~~~
\begin{subfigure}{.45\textwidth}
\begin{center}
\begin{lstlisting}[basicstyle=\ttfamily\tiny, frame=single, 
%caption={Comparison function which ignores the callee}, label=lst:fc_ign_callee, 
language=none,
numbers=left, stepnumber=1,
escapechar=@]
inst_post(@$\mathtt{?POST(f_c)}$@, @$\mathtt{f_{def}}$@, funs, n) =
    @$\mathtt{f'_{name}}$@ = get_free_name()
    @$\mathtt{f_{name}}$@ = get_name(@$\mathtt{f_{def}}$@)
    @$\mathtt{nfuns}$@ = {@$\mathtt{f_{name}}$@@$\mathtt{(\overline{fv_n}) \rightarrow}$@ edbc_lib:post(
        fun(@$\mathtt{fv_{res}}$@) @$\mathtt{\rightarrow}$@ 
            replace(@$\mathtt{body(f_c), \{?P(i) \mapsto fv_i, ?R \mapsto fv_{res}}\}$@)  
        end, 
        fun() @$\mathtt{\rightarrow}$@ @$\mathtt{f'_{name}}$@@$\mathtt{(\overline{fv_n})}$@ end).}
    return (rename_fun(@$\mathtt{f_{def}}$@, @$\mathtt{f'_{name}}$@), funs @$\cup$@ nfuns)
\end{lstlisting}
\vspace{-20pt}
\end{center}
\caption{Instrumentation of a \texttt{?POST} contract}
\label{fig:inst_post}
\end{subfigure}
~~~~~~~~~
\begin{subfigure}{.45\textwidth}
\begin{center}
\begin{lstlisting}[basicstyle=\ttfamily\tiny, frame=single, 
%caption={Comparison function which ignores the callee}, label=lst:fc_ign_callee, 
language=none,
numbers=left, stepnumber=1,
escapechar=@]
inst_put_info(@$\mathtt{f_{def}}$@, funs, n) =
    @$\mathtt{f'_{name}}$@ = get_free_name() @\label{fig:inst_pi_fresh}@ 
    @$\mathtt{f_{name}}$@ = get_name(@$\mathtt{f_{def}}$@)
    @$\mathtt{nfuns}$@ = {@$\mathtt{f_{name}}$@@$\mathtt{(\overline{fv_n}) \rightarrow}$@ @\label{fig:inst_pi_nf}@ 
        edbc_lib:put_info(@$\mathtt{f_{name}, \overline{fv_n}), f'_{name}(\overline{fv_n})}$@.} @\label{fig:inst_pi_call}@
    return (rename_fun(@$\mathtt{f_{def}}$@, @$\mathtt{f'_{name}}$@), funs @$\cup$@ nfuns) @\label{fig:inst_pi_rename}@ 
\end{lstlisting}
\vspace{-20pt}
\end{center}
\caption{Instrumentation to store information}
\label{fig:inst_pi}
\end{subfigure}
\caption{Instrumentation functions for contracts}
\label{fig:inst_funs}
\end{figure}

\begin{enumerate}

\item First of all, if the function has any contract associated (Figure \ref{fig:inst_alg}, line \ref{fig:inst_alg_pi}), then a instrumentation to store information of the call (function name, arguments and stack trace) is performed (Figure \ref{fig:inst_alg}, line \ref{fig:inst_alg_pi_2}). This instrumentation (Figure \ref{fig:inst_pi}) creates a new function which becomes the entry point of the original function (Figure \ref{fig:inst_pi}, line \ref{fig:inst_pi_nf}). This new function first stores the needed information and then calls to the original function (Figure \ref{fig:inst_pi}, line \ref{fig:inst_pi_call}) which has been renamed (Figure \ref{fig:inst_pi}, line \ref{fig:inst_pi_rename}) with a fresh name (Figure \ref{fig:inst_pi}, line \ref{fig:inst_pi_fresh}). In function \texttt{inst_put_info/3} and in the rest of instrumentation functions, variables \texttt{fv} represents fresh variables, function \texttt{get_free_name/0} returns a fresh function name, function \texttt{get_name/1} returns the function name of a given function definition, and function \texttt{rename_fun/2} renames a function definition.

\item Then, contracts of type \texttt{?DECREASES/1} (including \texttt{?SDECREASES/1}) are processed (Figure \ref{fig:inst_alg}, line \ref{fig:inst_alg_decr}). This instrumentation (Figure \ref{fig:inst_decr}) creates a function which checks if the size of its parameters have decreased between recursive calls (Figure \ref{fig:inst_decr}, line \ref{fig:inst_decr_new}). This new function ($\mathtt{f'_{name}}$) has two parameters, a list with the \texttt{o} arguments declared as decreasing of the previous call ($\mathtt{\overline{fv'_o}}$) and all the arguments of the next call  ($\mathtt{\overline{fv_n}}$). Its body is a single call to the \texttt{EDBC}'s function \texttt{decrease_check/3}, which receives the value of the decreasing arguments in the previous call ($\mathtt{\overline{fv'_o}}$) and in the next call ($\mathtt{ \overline{fv_n}|_{\overline{?P_o}}}$\footnote{Symbol $\mathtt{\overline{e_n} |_{\overline{?P_m}}}$ represents the selection of expressions in $\mathtt{\overline{e_n}}$ according to a given list of parameter numbers  $\mathtt{{\overline{?P_m}}}$. For instance, $\mathtt{[a,b,c,d] |_{[?P(1), ?P(3)]}}$ is equal to $\mathtt{[a,c]}$.}), and a delayed call to the contracted function's entry point. Using this information, function \texttt{decrease_check/3} compares previous and next values. If they are being decreased, it runs the delayed call, and if they are not, it raises a contract-violation error. During the instrumentation the original function is also modified by replacing all the recursive calls to calls to the new created function (Figure \ref{fig:inst_decr}, lines \ref{fig:inst_decr_replace1}-\ref{fig:inst_decr_replace2}). Function \texttt{replace/2} applies a substitution to an expression or a list of expressions. Note that, due to this instrumentation and the previous one, we have changed the call cycle of a recursive call from $\mathtt{f_{ori} \rightarrow f_{ori}}$ to $\mathtt{f_{si} \rightarrow f_{ori}  \rightarrow f_{dc}  \rightarrow f_{si}}$, where  $\mathtt{f_{ori}}$ is the original function, $\mathtt{f_{si}}$ is the function that stores the call information, and $\mathtt{f_{dc}}$ is the function that checks the decreasing of arguments.

\item Finally, the rest of the contracts are processed distinguishing among those contracts of type \texttt{?PRE} (Figure \ref{fig:inst_pre}) and those contracts of type \texttt{?POST} (Figure \ref{fig:inst_post}). All the contracts except \texttt{?DECREASE} can be generalized to one of of these two types of contract. Of course, each contract has its particularities (explained below), however these particularities do not have any effect in the instrumentation process. Both instrumentations create a function whose body is a call to a \texttt{EDBC}'s function (\texttt{pre/2} or \texttt{post/2}) that checks whether the pre- or the postcondition is held. This function receives two parameters: (1) the function which checks the conditions (named \emph{condition-checker function}) replacing in its body the parameter/result macros by their correspondent variables, and (2) a delayed call that, in the case of \texttt{?PRE} contracts, is run only in case the condition holds (a contract-violation is raised otherwise) or that, in the case of \texttt{?POST} contracts, is run before  contract checking and if the condition holds its result is returned (a contract-violation is raised otherwise). The original function is simply renamed like in the instrumentation which stores the call information. Therefore, when we call a function $\mathtt{f_{ori}}$ which defines pre- or a postcondition contracts, the chain of calls becomes $\mathtt{f_{si} \rightarrow f_{pre/post}*  \rightarrow f_{ori}}$, where  $\mathtt{f_{pre/post}}*$ are a number of functions (maybe none) introduced by \texttt{?PRE}/\texttt{?POST} contracts. In the case of a recursive function which defines a \texttt{?DECREASE} contract, the call chain would be $\mathtt{f_{si} \rightarrow f_{pre/post}*  \rightarrow f_{ori} \rightarrow f_{dc} \rightarrow f_{si}}$. Note that most of the helper functions have a call as its last expression enabling, in this way, last call optimizations. In fact, the only exception is the functions generated for postconditions. These functions should be stacked until its internal call is completely evaluated.

\end{enumerate}

In the rest of the section we explain the particularities of each contract type. Contracts of type \texttt{?DECREASE/1}, \texttt{?PRE/1} and \texttt{?POST/1} have not any particularity, i.e. they work as explained in the instrumentation process. The rest of contract types are implemented as described in the following.

\begin{itemize}

\item \textbf{Execution-time contracts}. These contracts are instrumented as \texttt{?PRE/1} contracts because the result, i.e. the parameter of the condition-checker functions of \texttt{?POST/1} contracts, is not needed. In fact, the evaluation of the condition-checker function does not return a boolean, but a number which represents the expected time. Then, the delayed call is run in the same process (\texttt{?EXPECTED_TIME/1}) or in a separate one (\texttt{?TIMEOUT/1}). Finally, according to the time needed to run the call, either the result is returned or a contract-violation error is raised (stopping also the evaluation in the case of \texttt{?TIMEOUT/1}).

\item \textbf{Purity contracts}. They are also implemented as \texttt{?PRE/1} contracts for the same reason as execution-time contracts. In this case, there is not any condition-checker function, so a dummy one is used. In order to check these contracts, we trace all the calls performed during the evaluation of the delayed call as well as receive/send events. The tracing process is performed using the BIF \texttt{erlang:trace/3}. A function call is considered impure if during its evaluation is performed a send, a receive or a call to an impure BIF (checked using the function \texttt{erl_bifs:is_pure/3}). 

\item \textbf{Invariant contracts}. These types of contracts are internally translated to \texttt{?POST/1} contracts and attached to functions which can change the state of an Erlang behaviour, e.g. \texttt{code_change/3},  \texttt{handle_call/3}, \texttt{init/1}, etc. Instead of the function result, the behaviour state is used to check whether the synthetized postcondition holds.

\item \textbf{Type contracts}. For checking the validity of the values according to the expected types we use the \texttt{Sheriff} \cite{Sheriff} type checker. \texttt{Sheriff} is run by calling the function \texttt{sheriff:check/2} which checks whether a given value belongs to a given type. We have gone a step forward making the type checking completely transparent for users and reusing their already-defined type contracts, i.e. their \texttt{spec}s. A \texttt{spec} implicitly defines a condition for the parameters and  a condition for the result. Therefore, a \texttt{spec} is translated to both a \texttt{?PRE/1} and a \texttt{?POST/1} contract which internally call to \texttt{Sheriff} and decide form its result whether to continue the evaluation or to raise a contract-violation error with attached details about the violator value and its expected type.

\end{itemize}

Finally, all these instrumentation processes have an effect in the total execution time that cannot be accepted in a final release. Therefore, after the code has been verified the code should return to its previous form, i.e. without instrumentation. For this reason, we have defined an easy mechanism to disable the instrumentation, e.g. with just a compilation flag.

\section{Concurrent contracts}
\label{sec:gen_server_cpre}

In this section, we explain the contracts that can only be used for concurrent evaluations. Although the contracts presented in Section \ref{sec:contracts} can still be really useful in this context, there are some scenarios where some special control is needed. This is where the contracts shown in this section shine. The section opens with the details of our approach and the contracts provided. The rest of the section proposes some scenarios where an implementation of these ideas can be applied to solve different kinds of problems. 

%

\subsection{\texttt{gen_server} with contracts}
\label{sec:gen_server_cpre}

We have extended the \texttt{gen_server} behaviour by adding the callback function \texttt{cpre/3} which allows users to decide whether the server is ready or not to serve a given request. The rest of the \texttt{gen_server}'s callbacks are not modified. The three parameters of function \texttt{cpre/3} are 1) the request, 2) the \emph{from of the request}\footnote{The \emph{from of the request} has the same form than in the \texttt{handle_call/3} callback, i.e. a tuple \texttt{\{Pid,Tag\}}, where \texttt{Pid} is the pid of the client and \texttt{Tag} is a unique tag.} and 3) the current server state. The function \texttt{cpre/3} should evaluate to a tuple with two elements. The first tuple's element is a boolean value which indicates if the given request can be served. The second tuple's element is the new server state which can be modified, e.g. by adding some control on the delayed requests. 

The \texttt{gen_server_cpre} behaviour behaves exactly in the same way than the \texttt{gen_server} behaviour but with a particularity. Each time the server receives a synchronous request, it calls to \texttt{cpre/3} callback before calling the actual \texttt{gen_server} callback, i.e. \texttt{handle_call/3}. Then, according to the value of the first element of the tuple that \texttt{cpre/3} returns, either the request is actually performed (when the value is \texttt{true}) or it is queued to be served later  (when the value is \texttt{false}). In both cases, the server state is updated with the value returned in the second element of the tuple. Thus, a \texttt{cpre/3} callback can be seen as a kind of contract for concurrent environments. 

\texttt{EDBC} includes two implementations of the \texttt{gen_server_cpre} behaviour, each one treats the queued requests in a different way. The simplest one is \texttt{gen_server_cpre} behaviour that resends to itself an unserveable request, i.e. a request for which function \texttt{cpre/3} returns \texttt{\{false, \ldots\}}). Since mailboxes in Erlang are FIFO, the posponed request will be the last request in the queue of pending requests. This can be considered as unfair because, once the server's state is ready to serve the posponed request, it could serve requests that arrived later instead. Therefore, \texttt{EDBC} also provides a fairer version of the \texttt{gen_server_cpre} behaviour. In this version, three queues are used to control that older requests are served first: $\mathtt{queue_{current}}$, $\mathtt{queue_{old}}$,  and $\mathtt{queue_{new}}$. Each time the server is ready to listen for new requests, the $\mathtt{queue_{current}}$ is inspected. If it is empty, then the server proceeds as usual, i.e. by receiving a request from its mailbox. Otherwise, if it is not empty, a request from $\mathtt{queue_{current}}$ is served. Consequently, the taken request is removed from $\mathtt{queue_{current}}$. The queues are also modified by adding requests to $\mathtt{queue_{old}}$ and $\mathtt{queue_{new}}$. This is done when function \texttt{cpre/3} returns \texttt{\{false, \ldots\}}. Depending on the origin of the request it is added to $\mathtt{queue_{old}}$ (when it comes from $\mathtt{queue_{current}}$) or to $\mathtt{queue_{new}}$ (when it comes from the mailbox). Finally, each time a request is completely served, the server's state could have been modified. A modification in the server's state can enable posponed requests. Therefore, each time the server's state is modified, $\mathtt{queue_{current}}$ is rebuilt as follows:  $\mathtt{queue_{old}}~+~\mathtt{queue_{current}}~+~\mathtt{queue_{new}}$.

\subsection{Selective receives}
\label{sec:sel_recv}

We have found several posts where some limitations of the \texttt{gen_server} implementation are discussed. Most of them are related on how to implement a server which has the ability to delay some requests. One of theses examples is the following question asked in stackoverflow.com.

\begin{center}\url{https://stackoverflow.com/questions/1290427/how-do-you-do-selective-receives-in-gen-servers}\end{center}

The questioner asked whether it was possible to implement a server which performs a selective receive while using a \texttt{gen_server} behaviour. None of the provided answers is giving an easy solution. Some of them suggest that the questioner should not use a \texttt{gen_server} for this, and directly implement a \emph{low-level} selective receive. Other answers propose to use \texttt{gen_server} but delay the requests \emph{manually}. This solution involves storing the request in the server state and returning a \texttt{no_reply} in the \texttt{handle_call/3}. Then, the request should be revised continually, until it can be served and use \texttt{gen_server:reply/2} to inform the client of the result. Our solution is closer to the last one, but all the management of the delayed requests is completely transparent for the user. 

%
%

\begin{wrapfigure}[9]{r}{0.55\textwidth}
\vspace{-30pt}
\begin{center}
\begin{lstlisting}[basicstyle=\ttfamily\scriptsize, frame=single, 
%caption={Comparison function which ignores the callee}, label=lst:fc_ign_callee, 
language=erlang,
numbers=left, stepnumber=1,escapechar=@]
handle_call(test, _From, _State) ->
  List = [0,1,2,3,4,5,6,7,8,9], @\label{fig:sel_recv_1}@
  lists:map(fun(N) -> spawn(fun() -> 
      gen_server:call(?MODULE, {result, N}) end) 
    end, lists:reverse(List)),
  {reply, ok, List}; @\label{fig:sel_recv_2}@
handle_call({result, N}, _From, [N|R]) ->
  io:format("result: " ++ integer_to_list(N) ++ "~n"),
  {reply, ok, R}.
\end{lstlisting}
\vspace{-20pt}
\end{center}
\caption{\texttt{handle_call/2} for selective receive}
\label{fig:sel_recv}
\vspace{-10pt}
\end{wrapfigure}

Figure \ref{fig:sel_recv} shows the function \texttt{handle_call/2} of the \texttt{gen_server} that the questioner provided to exemplify the problem. When the request \texttt{test} is served, it builds ten processes, each one performing a \texttt{\{result, N\}} request, with \texttt{N} ranging from 0 to 9. Additionally, the server state is defined as a list which also ranges from 0 to 9 (Figure \ref{fig:sel_recv}, lines \ref{fig:sel_recv_1} and \ref{fig:sel_recv_2}). The interesting part of the problem is how the \texttt{\{result, N\}} requests need to be served. The idea of the questioner is that the server should process the requests in the order defined by the state. For instance, the request \texttt{\{result, 0\}} can only be served when the head of the state's list is also \texttt{0}. However, there is a problem in this approach. The questioner explains it with the sentence: \emph{when none of the callback function clauses match a message, rather than putting the message back in the mailbox, it errors out}. Although this is the normal and the expected behaviour of a \texttt{gen_server}, the questioner thinks that some easy alternative should exists. However, as explained above, the solutions proposed in the thread are not satisfactory enough. 

\begin{wrapfigure}[5]{r}{0.5\textwidth}
\vspace{-30pt}
\begin{center}
\begin{lstlisting}[basicstyle=\ttfamily\scriptsize, frame=single, 
%caption={Comparison function which ignores the callee}, label=lst:fc_ign_callee, 
language=erlang,
numbers=left, stepnumber=1,escapechar=@]
cpre(test, _, State) -> {true, State};
cpre({result, N}, _, [N|R]) -> {true, [N|R]};
cpre({result, N}, _, State) -> {false, State}.
\end{lstlisting}
\vspace{-20pt}
\end{center}
\caption{\texttt{cpre/3} for selective receive}
\label{fig:sel_recv_cpres}
\vspace{-10pt}
\end{wrapfigure}

With the enhanced versions of the \texttt{gen_server} behaviour we propose in this paper, users can define conditions for each request by using function \texttt{cpre/3}. Figure \ref{fig:sel_recv_cpres} depicts a definition of the function \texttt{cpre/3} that solves the questioner's problem without needing to redefine function \texttt{handle_call/3} of Figure \ref{fig:sel_recv}. The first clause indicates to the \texttt{gen_server_cpre} server that the request \texttt{test} can be served always. Contrarily, \texttt{\{result, N\}} requests only can be served when \texttt{N} coincides with the first element of the server's state. Thus, with this simple code addition the questioner has a selective receive in a \texttt{gen_server} context.

\subsection{Readers-writers example}
\label{sec:rw}

\begin{wrapfigure}[12]{r}{0.6\textwidth}
\vspace{-30pt}
\begin{center}
\begin{lstlisting}[basicstyle=\ttfamily\scriptsize, frame=single, 
%caption={Comparison function which ignores the callee}, label=lst:fc_ign_callee, 
language=erlang,
numbers=left, stepnumber=1,escapechar=@]
handle_call(request_read, _, State) ->
  NState = State#state{readers = State#state.readers + 1},
  {reply, pass, NState};
handle_call(request_write, _, State) ->
  NState =  State#state{writer = true}},
  {reply, pass, NState}.

handle_cast(finish_read, State) ->  
  NState = State#state{readers = State#state.readers - 1},
  {noreply, NState};
handle_cast(finish_write, State) -> 
  NState = State#state{writer = false},
  {noreply, NState}.
\end{lstlisting}
\vspace{-20pt}
\end{center}
\caption{Readers-writers request handlers}
\label{fig:rw_handlers}
\vspace{-10pt}
\end{wrapfigure}

In this section we define a simple server that implements the readers-writers problem. We start introducing an implementation of the problem using the standard \texttt{gen_server} behaviour. The server's state is a record defined as \texttt{-record(state, {readers = 0, writer = false})}. The requests that it can handle are four: \texttt{request_read}, \texttt{request_write}, \texttt{finish_read} and \texttt{finish_write}. The first two requests are synchronous (because clients need to wait for a confirmation) while the latter two are asynchronous (clients do not need confirmation). Figure \ref{fig:rw_handlers} shows the handlers for these requests. They basically increase/decrease the counter \texttt{readers} or switch on/off the flag \texttt{writer}.

\begin{wrapfigure}[6]{r}{0.6\textwidth}
\vspace{-40pt}
\begin{center}
\begin{lstlisting}[basicstyle=\ttfamily\scriptsize, frame=single, 
%caption={Comparison function which ignores the callee}, label=lst:fc_ign_callee, 
language=erlang,
numbers=left, stepnumber=1,escapechar=@]
?INVARIANT(fun invariant/1).

invariant(#state{ readers = Readers, writer = Writer}) -> 
    is_integer(Readers) andalso Readers >= 0 
    andalso is_boolean(Writer)
    andalso ((not Writer) orelse Readers == 0).
\end{lstlisting}
\vspace{-20pt}
\end{center}
\caption{Readers-writers invariant definition}
\label{fig:rw_invariant}
\vspace{-10pt}
\end{wrapfigure}

Having defined all these components, we can already run the readers-writer server. It will start serving requests successfully without any noticeable issue. However, the result in the shared resource would be a mess, mainly because we are forgetting a really important piece in this problem: its invariant, i.e. $!writer \vee readers = 0$.
We can define an invariant for the readers-writers server by using the macro \texttt{?INVARIANT/1} introduced in Section \ref{sec:contracts}. Figure \ref{fig:rw_invariant} shows how the macro is used and the helper function which actually checks the invariant. Apart from the standard invariant, i.e. \texttt{(not Writer) orelse Readers == 0}, the function also checks that the state's field \texttt{readers} is a positive integer and that the state's field \texttt{writer} is a boolean value.

If we rerun the server with the invariant defined, we obtain some feedback on whether the server is behaving as expected. In this case, the server is not a correct implementation of the problem. Therefore, an error should be raised due to the violation of the invariant. An example of these errors is shown in Figure \ref{fig:rw_fail_invariant}. The error is indicating that the server's state was \texttt{\{state,0,true\}} when the server processed a \texttt{request_read} which led to the new state \texttt{\{state,1,true\}} which clearly violates the defined invariant. The information provided by the error report can be improved by returning a tuple \texttt{\{false, Reason\}} in the invariant function, where \texttt{Reason} is a string to be shown in this contract-violation report after the generic message.

\begin{wrapfigure}[10]{r}{0.45\textwidth}
\vspace{-35pt}
\begin{center}
\begin{lstlisting}[basicstyle=\ttfamily\scriptsize, frame=single,
% caption={\texttt{SecEr} reports that no discrepancies exist}, label=lst:happycorrect, 
language=none,escapechar=@]
=ERROR REPORT==== 
** Generic server readers_writers terminating 
** Last message in was request_read
** When Server state == {state,0,true}
** Reason for termination == 
** {{"The invariant does not hold.",
Last call: readers_writers:handle_call(
request_read, @\ldots@, {state,0,true}). 
Result: {reply, pass,{state,1,true}}",
 [{readers_writers,handle_call,3,@\ldots@},
    @\ldots@]}, @\ldots@}
\end{lstlisting}
\vspace{-20pt}
\end{center}
\caption{Failing invariant report}
\label{fig:rw_fail_invariant}
\vspace{-10pt}
\end{wrapfigure}

In order to correctly implement this problem, we use function \texttt{cpre/3} to control when a request can be served or not. Figure \ref{fig:rw_cpres} shows a function \texttt{cpre/3} which makes the server's behaviour correct and avoids violations of the invariant. It enables \texttt{request_read} requests as long as the flag \texttt{writer} is switched off. Similarly, the \texttt{request_write} requests also require the flag \texttt{writer} to be switched off and also the counter \texttt{readers} to be 0. If we rerun now the server, no more errors due to invariant violations will be raised.

Although this implementation is already correct, it is unfair for writers as they have less chances to access the shared resource. We have several alternative implementations in the GitHub repository\footnote{\url{https://github.com/tamarit/edbc/tree/master/examples/readers_writers}}, some of them are fairer than the solution presented above. All the solutions are implemented using the behaviour \texttt{gen_server_cpre}.

\begin{wrapfigure}[8]{r}{0.57\textwidth}
\vspace{-40pt}
\begin{center}
\begin{lstlisting}[basicstyle=\ttfamily\scriptsize, frame=single, 
%caption={Comparison function which ignores the callee}, label=lst:fc_ign_callee, 
language=erlang,
numbers=left, stepnumber=1,escapechar=@]
cpre(request_read, _, State = #state{writer = false}) ->
  {true, State};
cpre(request_read, _, State) ->
  {false, State};
cpre(request_write, _, 
    State = #state{writer = false, readers = 0}) ->
  {true, State};
cpre(request_write, _, State) ->
  {false, State}.
\end{lstlisting}
\vspace{-20pt}
\end{center}
\caption{Readers-writers \texttt{cpre/3} definition}
\label{fig:rw_cpres}
\vspace{-10pt}
\end{wrapfigure}

The examples presented in Sections \ref{sec:sel_recv} and \ref{sec:rw} are some of the examples available at the the GitHub repository\footnote{\url{https://github.com/tamarit/edbc/tree/master/examples}}. In the examples directory there are implementations of semaphores, single-lane bridges and other classical problems in concurrency.

\section{Related Work}
\label{sec:related}


Our contracts are similar to the ones defined in \cite{AH12}, where the function specifications are written in the same language, i.e. Curry, so they are executable. Being executable enables their usage as prototypes when the real implementation is not provided. Their contracts are functions in the source code instead of macros, so it is not clear whether they could be removed in the final release. There is not any mention about whether their contracts are integrated with a documentation tool like our contracts are with EDoc. Moreover, they only allow to define basic precondition and postcondition contracts, while we are providing alternative ones like purity or time contracts. Finally, our contracts for concurrent environments have a really different approach. 

The work in \cite{PS10} presents a static analysis which infers whether a function is pure or not. Since they are working in a static context and ours is dynamic, the purity checking is performed in completely different ways in each work. However, we can benefit from their results by, for instance, avoiding to execute functions that are already known to be impure, reporting earlier to the user a purity-contract violation. In the same way, our system can be used by them to check the validity of their statically-inferred results.

The type contract language for Erlang \cite{JLS07} allows to specify the intended behavior of functions. Their usage is twofold: 1) as a documentation in the source code which is also used to generate EDoc, and 2) to refine the analysis such as the one that performs Dialyzer. The contract language allows for singleton types, unions of existing types and the definition of new types. However, these types and function specifications do not guarantee type safety. This guarantee comes with Erlang which incorporates a strong typing that reports type errors during runtime. Although it is a powerful and useful analysis, strong typing usually detects an unexpected behaviour too far from its source. Therefore, when debugging a program, these type contracts can be really useful to report a type-contract violation earlier. This was the motivation to incorporate them in our system.

The contracts proposed for the concurrent environments follow the same philosophy as the specifications defined in \cite{HMCM09,FMAH16}. Indeed, our function \texttt{cpre/3} takes its name from these works. Although these works were more focused in enabling the use of formal methods in order to verify nontrivial properties of realistic systems, in this paper we demonstrate that they can be also really useful for runtime verification and for the management of temporally-unservable requests.

Dafny \cite{Lei10} is a language which allows to define invariants and contracts in their programs. The main difference between their approach and ours is that their contracts are not checked during runtime, but during compile-time. Additionally, as we have explained in Section \ref{sec:contracts_exa}, we can replicate the same type of contracts in \texttt{EDBC}. However, our approach does not need an extra syntax or functionality to define complex contracts as Dafny does. 

The aspect-oriented approach for Erlang (\texttt{eAOP}) presented in \cite{CFAI17} shares some similarities with our work. eAOP allows the instrumentation of a program with a user-defined tracing protocol (at an expression level). This is able to report events to a monitor (asynchronous) as well as to force some part of the code to block waiting for information from the monitor (synchronicity). Our system could be used to a similar purpose but only at the function level. Additionally, thanks to the functionality freedom  allowed in our contracts, our system enables the definition of synchronization operations at the user-defined contracts. More complex modifications of our system, such as the ones done in \cite{LS05}, can transform our work into a complete aspect-oriented system. 

Also in Erlang, the work \cite{CFG11} defines a runtime monitoring tool which helps to detect such messages that do not match with a given specification. These specifications are defined through an automaton, which requires an extra knowledge from the user in both, the syntax of the specification and in the whole system operation. We propose a clear and easy way to define the conditions to accept or not a request without needing any special input from users.

Finally, JErlang \cite{PE10} enables joins in Erlang. The authors reach their goal by increasing the syntax of the receive patterns so they could express matching of multiple subsequent messages. Although our goal and theirs are different, both approaches can simplify the way programmers solve similar kind of problems. Indeed, we could simulate joins by adding a forth parameter to the function \texttt{cpre/3}. This additional parameter would represent the unserved/pending requests. When the last request of a user-defined sequence (\emph{join}) was received, the pending requests would be examined to check whether the required join could be served. A similar modification would be needed in \texttt{handle_call/3} so the pending requests should be duly served by using \texttt{gen_server:reply/2}.

\section{Conclusions}
\label{sec:conclusions}

We have proposed different types of contracts for general Erlang which help programmers to define otherwise-lost knowledge of their programs' expected behaviour. These contracts are meant to be checked during runtime, although its checking can be disabled for the final release. Additionally, automatic documentation can be generated from them. Our contracts use plain Erlang, without the need of defining new syntax or supporting libraries. We have also defined contracts for concurrent systems. These contracts solve scenarios like requests that should be served or not according to the server's state. The simplicity of the approach and its integration in the Erlang behaviours ease the programming of systems using the proposed features. 

There are multiple extensions of this work. For instance, by extending contracts to return a default value instead of an error when a contract is violated. A more general  is to leave it up to the users how the system should react when a contract is violated. We also need a way to express errors, e.g. to define that we expect that when certain argument is wrong the called function returns an error. We could use tracing instead of transforming the program to enable recursive paths visiting more than one function. The decrease contracts can be also extended to receive the comparison function to be used when deciding that certain parameter is being decreased. 

We can use our concurrent-contract approach to control starvation of systems. The idea is to use a mapping that goes from types of request to waiting requests, and that represents the fact that a delayed request is waiting for a concrete event to occur. The invariant can be used then to control starvation. Another extension consists in that the preconditions and postconditions used inside an Erlang behaviour could make the client of the resource to fail instead of the resource itself. 

In a more general view, we can extend our system to automatically derivate EUnit or property tests from the contracts. In a similar way, contracts could be used in a property-based testing environment to define, for instance, that all the inputs of a function always hold a given property. With respect to invariants, we could also attach them to the function \texttt{spawn} in order to enable the definition of properties such as: the number of pending messages for this process cannot be greater than 1. Finally, we are trying to establish a relation between liquid types and our approach as we have found partial similarities.

%
%
%
%
%
%
%

\bibliography{\biblio{biblio}}
\bibliographystyle{abbrv}

\end{document}

\newpage
\noindent \underline{Note for the reviewers:} The following appendix has been only included to ease the reviewing process, and it will not be part of the final paper. In case of acceptance, this appendix will be published as a technical report so that the interested reader will have public access to it.

\appendix
\label{appendix}

\input{./Secciones/Pruebas.tex}

\end{document}